\documentclass{Interspeech}

\interspeechcameraready
\title{Assessing the Impact of Anisotropy in Neural Representations of Speech: A Case Study on Keyword Spotting}

\author[affiliation={1}]{Clara}{Rosina Fernandez}
\author[affiliation={1}]{Séverine}{Guillaume}
\author[affiliation={2}]{Guillaume}{Wisniewski}

\affiliation{}{LACITO, CNRS, Université Sorbonne Nouvelle}{F-94800 Villejuif, France}\affiliation{}{LLF, CNRS, Université Paris-Cité}{F-75013 Paris, France}
\email{c.rosinafernandez@gmail.com, severine.guillaume@cnrs.fr, guillaume.wisniewski@u-paris.fr}

\usepackage{amsmath,amssymb}
\usepackage{subcaption}
\usepackage{multirow}
\usepackage{makecell}

\begin{document}

\maketitle

\begin{abstract}    
Pretrained speech representations like \texttt{wav2vec2} and \texttt{HuBERT} exhibit strong anisotropy, leading to high similarity between random embeddings. While widely observed, the impact of this property on downstream tasks remains unclear. This work evaluates anisotropy in keyword spotting for computational documentary linguistics. Using Dynamic Time Warping, we show that despite anisotropy, \texttt{wav2vec2} similarity measures effectively identify words without transcription. Our results highlight the robustness of these representations, which capture phonetic structures and generalize across speakers. Our results underscore the importance of pretraining in learning rich and invariant speech representations.
\end{abstract}

\section{Introduction}

Contextually pretrained speech representations, such as those extracted by \texttt{wav2vec2}~\cite{baevski20wav2vec2} or \texttt{HuBERT}~\cite{hsu21hubert} from raw data, play a crucial role in many applications. However, several studies \cite{gao2018representation,timkey-van-schijndel-2021-bark,rajaee-pilehvar-2021-fine-tuning,ait-saada-nadif-2023-anisotropy,godey-etal-2024-anisotropy} on transformers analyzing these representations have highlighted their strong anisotropy: The distribution of vectors in the representation space is highly uneven, with certain directions being more prominent than others and, as consequence, the similarity between two embeddings chosen randomly is often close to 1~\cite{ethayarajh-2019-contextual}. 

Intuitively, this property imposes a significant limitation on the expressiveness of models and constrains their usability. Yet, while anisotropy has been widely observed and various methods have been proposed to mitigate its effects, to our knowledge, no study has assessed its actual impact on downstream tasks: The results reported in the related works are generally limited to measuring the average distribution of distances between representations of two frames without taking into account the way in which these representations will be used. The primary objective of this work is to evaluate this impact taking the keyword spotting as a use-case. This application is directly motivated by a use-case in computational documentary linguistics: Our goal is to help linguists in their work of documenting and analyzing languages by letting them easily find all the occurrences of a word in their corpus. More precisely, we use the classical DTW algorithm to find all occurrences of an audio “query” in a corpus by directly comparing a representation of the query with the representations of the recordings in the corpus. Our experiments, described in Section~\ref{sec:exp}, show that, contrary to what the anisotropy of representations might suggest, it is possible to directly use similarity measures between \texttt{wav2vec2} representations to identify words or word groups within speech corpora without requiring transcription.

This result incidentally evaluated one particularly interesting aspect of \texttt{wav2vec2} representations, namely their robustness: Even if they are trained with a \emph{reconstruction loss} (where the model learns to predict missing or corrupted parts of the input signal) that measure their capacity to capture all aspects of the input signal, \texttt{wav2vec2} constructs abstract representations that capture high-level phonetic structures. As a result, the distance between two occurrences of the same word, even when spoken by different persons, remains small, highlighting the model’s ability to generalize across speakers. 

The rest of this article is organized as follows. In Section~\ref{sec:anisotropy}, we briefly review the literature on the anisotropy of transformer representations and show that \texttt{wav2vec2} representations suffer from the same issue. Then, in Section~\ref{sec:exp}, we describe our keyword spotting experiments designed to assess the impact of anisotropy on downstream tasks and characterize the space of audio representations constructed by this model.

\section{Anisotropy in Neural Representations \label{sec:anisotropy}}

The study of the geometric properties of representation spaces, automatically constructed by neural networks, has gained significant attention in recent years~\cite{peters-etal-2018-dissecting}. Interestingly, research has shown that these representations are not uniformly distributed within the vector space. Instead, they tend to concentrate within a narrow cone ---~a counterintuitive phenomenon given the high-dimensional nature of these spaces. This is particularly surprising considering the success of these representations in various applications, suggesting that, whether they encode word representations or direct audio representations, they capture rich information about both the phonological and morphosyntactic properties of words.

In the context of text representations, this anisotropy has been observed across various neural models. \cite{ethayarajh-2019-contextual} demonstrated that early Transformer-based models (such as \texttt{BERT} and \texttt{GPT-2}) exhibit strong anisotropy. \cite{godey-etal-2024-anisotropy} has extended these first observations to other modality (and, in particular, representation of speech) and more recent models (such as \texttt{T5}), suggesting that anisotropy is actually inherent to Transformers-based models. 

To quantify this anisotropy, \cite{timkey-van-schijndel-2021-bark} proposed measuring it using the expected cosine similarity between randomly selected word representations in a given corpus. Formally,the anisotropy measure $\mathcal{A}$ is defined as:
\begin{equation}
\mathcal{A} = \mathbb{E}_{i \neq j}\left( 1 - \cos \left(\mathbf{x}_i, \mathbf{x}_j\right)\right)
\end{equation}
where $\left(\mathbf{x}_i, \mathbf{x}_j\right)$ are word or audio representations randomly drawn from the corpus, and $\mathbb{E}[\cdot]$ denotes the expectation over all such pairs. Intuitively, if this measure is close to 1, the representations are highly anisotropic, meaning they are concentrated in a restricted region of the vector space. Conversely, if the value approaches 0, it indicates that the representations are more evenly spread across all possible directions. A key insight from \cite{timkey-van-schijndel-2021-bark} is that this anisotropy primarily stems from the presence of a small number of rogue dimensions in contextual embeddings. These dimensions exhibit exceptionally large magnitudes and high variance, dominating cosine similarity computations and skewing the overall distribution of representations. 

To illustrate this anisotropy of representations, we estimated the $\mathcal{A}$ parameter of \texttt{wav2vec2}, following methods used in previous studies on neural representations anisotropy. We used \texttt{XLSR-53}~\cite{conneau21unsupervised}, a pretrained \texttt{wav2vec2}  model, which was trained on 56,000 hours of multilingual audio data across 53~languages. We randomly selected 1,000~recordings from the Spanish subset of the \texttt{CommonVoice} corpus~\cite{conneau21unsupervised}. Each recording was processed using \texttt{XLSR-53}, resulting in a sequence of $1,024$-dimensional vectors for each of the 24 layers of the model, where each vector represents 1/49 of a second of speech. We then randomly sampled 1,000~pairs of vectors from this corpus and computed their cosine similarity.

\begin{figure*}
\includegraphics[width=.9\textwidth]{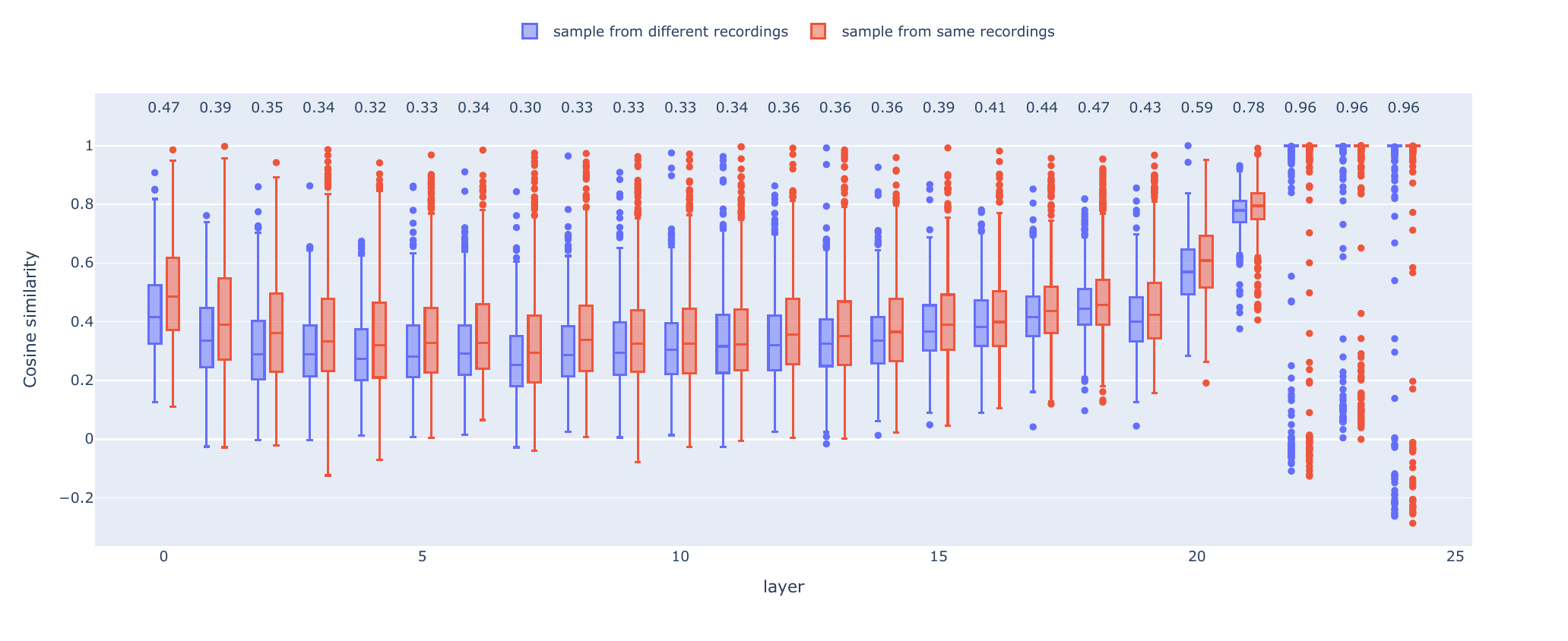}
\centering
\caption{Distribution of cosine similarities across layers, measured between two audio embeddings determined by \texttt{XLSR-53}. The anisotropy measure $\mathcal{A}$ for each layer is also reported.\label{fig:anistropy}}
\end{figure*}

The value of $\mathcal{A}$ measured on this sample is 0.46, confirming the anisotropy of \texttt{wav2vec2} models. To further refine this observation, we have plotted the distribution of cosine similarities in Figure~\ref{fig:anistropy}, distinguishing between cases where the representations come from the same recording and those from different recordings. While these distributions are clearly not centered on 0 (with almost no negative similarities!), the situation appears more complex than initially thought: anisotropy is particularly strong in the last layers of the model. Surprisingly, the distributions of similarities between representations from the same and different recordings are similar, with the median and IQR being of the same order of magnitude. 

To demonstrate the presence of rogue dimensions in the representations, we report in Table~\ref{tab:rogue} the two dimensions with the highest mean values, along with the median of mean values across all dimensions, to analyze the value distribution. Results, in Table~\ref{tab:rogue}, reveal the systematic presence of rogue dimensions, characterized by disproportionately high mean values. In every layer (except layer 0, which corresponds to non-contextualized representations used as Transformer input), the highest-mean dimension exceeds the second-highest by several orders of magnitude, highlighting extreme activation imbalance. This effect is particularly pronounced in layer 22 and 23, where the dominant dimension shows an exceptionally large deviation.

\begin{table}
{\footnotesize
\begin{tabular}{ccccc}
\toprule
layer & \makecell{max. mean \\ across dim.} & \makecell{std across \\  max dim.} & second max. & median of means \\
\midrule
0 & 18   & 18 & 17 & 0.7 \\
1 & 43   & 45 & 16 & 1.0 \\
2 & 40   & 59 & 14 & 1.0 \\
3 & 45   & 65 & 15 & 1.0 \\
4 & 51   & 107 & 13 & 0.9 \\
5 & 59   & 275 & 14 & 0.9 \\
6 & 65   & 1,230 & 14 & 0.9 \\
7 & 68   & 3,020 & 11 & 0.9 \\
8 & 75   & 3,740 & 11 & 0.9 \\
9 & 85   & 7,988 & 13 & 0.9 \\
10 & 86  & 7,137 & 15 & 0.9 \\
11 & 88  & 8,748 & 16 & 0.9 \\
12 & 90  & 9,116 & 15 & 0.9 \\
13 & 88  & 5,808 & 14 & 0.9 \\
14 & 82  & 4,599 & 11 & 0.9 \\
15 & 91  & 4,256 & 10 & 0.9 \\
16 & 97  & 2,485 & 10 & 0.9 \\
17 & 108 & 2,450 & 11 & 0.9 \\
18 & 116 & 1,084 & 13 & 0.9 \\
19 & 129 & 1,189 & 14 & 0.9 \\
20 & 184 & 2,534 & 55 & 0.7 \\
21 & 286 & 2,727 & 29 & 0.5 \\
22 & 27,151 & 129,132,920 & 7,948 & -458.7 \\
23 & 28,145 & 126,904,352 & 8,154 & -473.1 \\
24 & 2 & 0 & 1 & -0.0 \\
\bottomrule
\end{tabular}
}
\caption{Demonstration of rogue dimensions in \texttt{XLSR-53} representations: For each layer, we report the two highest mean values along representation dimensions, along with the median of the means across dimensions and the standard deviation of values in the largest dimension. All values are rounded to the nearest integer.\label{tab:rogue}}
\end{table}

All these observations suggest that the similarity values are difficult to interpret as they remain close regardless of whether the utterances were produced by the same speaker, by individuals with “similar” voices, or even by different speakers discussing unrelated topics.

\section{Is Anistropy Harmfull?\label{sec:exp}}

\paragraph*{Context} To assess the impact of anisotropy we consider the keyword spotting task. Keyword spotting refers to the task of detecting specific words or phrases within an audio corpus without requiring full transcription. This task has numerous applications, including voice search, spoken document retrieval, command recognition in voice assistants, and speaker-independent speech analytics. In this work, we focus on keyword spotting in the context of computational documentary linguistics, where the goal is to assist linguists in retrieving all occurrences of a given word or phrase within a speech corpus. As described by \cite{macaire-etal-2022-automatic}, this capability is particularly valuable for language documentation, enabling researchers to study phonetic variations, lexical patterns, and language usage in low-resource or endangered languages. By efficiently locating words across recordings,\footnote{In this work, we assume that the corpus is segmented into phrases and that the search query consists of one or multiple words.} linguists can analyze their distribution and usage without requiring full transcription, which is often impractical in under-documented languages, as, for many of them, linguists do not have the resources to learn a transcription system (whether it is annotated data, skills or computing infrastructure).

The most natural approach to solving this task is as follows: given a set of $N$ audio recordings, each represented as a sequence of feature vectors, and a query also represented as a sequence of $m$ feature vectors, the objective is to rank the recordings from most to least similar to the query. This task reduces to measuring the similarity between two sequences of vectors, where one sequence (the query) may be a subsequence of the other (the target recording). More formally, we seek to find the best alignment between the query and segments of the target audio representations.

A classical approach for solving this problem is Subsequence Dynamic Time Warping \cite{muller15fundamentals}. This algorithm operates on an $n \times m$ similarity matrix, where $n$ is the number of frames in the considered recording and each element represents the similarity between a frame of the recording and a frame of the query. Subsequence DTW efficiently finds the optimal alignment between the query and a subsequence of the target by allowing non-linear time alignment and handling variations in speech rate and pronunciation.

In addition to its practical interest, this task allows us to measure the impact of anisotropy. The calculated alignment, and therefore the similarity between a query and a record, directly depends on the similarity between the vectors representing frames. If all similarities are equal, the alignment determined by DTW will be meaningless, making it unable to distinguish records that contain the word from those that do not. Conversely, if our approach successfully identifies sentences containing the query words with high precision, then DTW similarities ---~despite relying on subtle variations~--- will still be relevant. In this case, we can conclude that anisotropy has only a limited impact: even if the absolute values of similarities are difficult to interpret (since they are all very close), they can still be used to identify linguistically similar records.


\paragraph*{Experimental Setup}
In our experiments, we utilized the Spanish portion of \texttt{CommonVoice}~\cite{ardila-etal-2020-common} and selected 30 target words. For each target word, we identified 10 sentences within the transcriptions that contained the target word. This resulted in a total of 300 queries, each representing different pronunciations of the same 30 words to ensure a variety of accents, contexts, and pitches. To validate the effectiveness of our representation, we ensured that each set included the recording from which the query was extracted. This allowed us to verify whether the representations could reliably identify the “source” recording. Additionally, we randomly selected 700 sentences that did not contain any of the target words. The objective of this experiment, as outlined in the previous section, is to determine whether the 10 sentences containing each query word can be correctly retrieved from the pool of 1,000 sentences.

To perform these experiments, we needed the audio segments corresponding to the query words. We used the Montreal Forced Aligner~\cite{mcauliffe17mfa} to identify which part of the audio signal corresponded to each word in the transcriptions. Based on this alignment, we constructed three different representations of each query word:
\begin{itemize}
\item MFCC Representation (Baseline): The aligned audio segment corresponding to the target word was extracted, and its representation was computed using 13 MFCCs with a 25\,ms slidding window;
\item \texttt{XLSR-53} Word Representation: The word's audio segment was extracted based on alignment information and encoded using \texttt{XLSR-53};
\item \texttt{XLSR-53} Contextual Representation: The entire audio utterance was first encoded using \texttt{XLSR-53}, and the word’s representation was obtained by extracting the vectors corresponding to the aligned word segment. This representation allows us to measure the influence of contextual information on the representation of words.
\end{itemize}

To retrieve relevant sentences for each query, we applied the Subsequence DTW algorithm, using cosine similarity as the base similarity metric. This method was used to rank the 1,000 sentences in our dataset for each query, and we verified whether the retrieved sentences contained the target word based on their transcriptions.\footnote{This evaluation provides easily interpretable \texttt{Precision@k} and \texttt{Recall@k} results while assessing the system's usability for linguistic documentation tasks (finding relevant records). A finer-grained evaluation—checking whether the word is accurately located in the record—is possible via DTW-based alignment, but we found it harder to interpret.}

To evaluate retrieval performance, we computed \texttt{Precision@k} and \texttt{Recall@k} for each query. These metrics are defined as follows: \texttt{Precision@k} represents the proportion of retrieved sentences in the top $k$ results that contain the target word, while \texttt{Recall@k} corresponds to the proportion of target-containing sentences that appear among the top $k$ retrieved results. Finally, we report both metrics averaged over documents (i.e., macro scores).

\begin{figure*}
    \centering
    \begin{subfigure}{0.45\textwidth}
    \includegraphics[width=\linewidth]{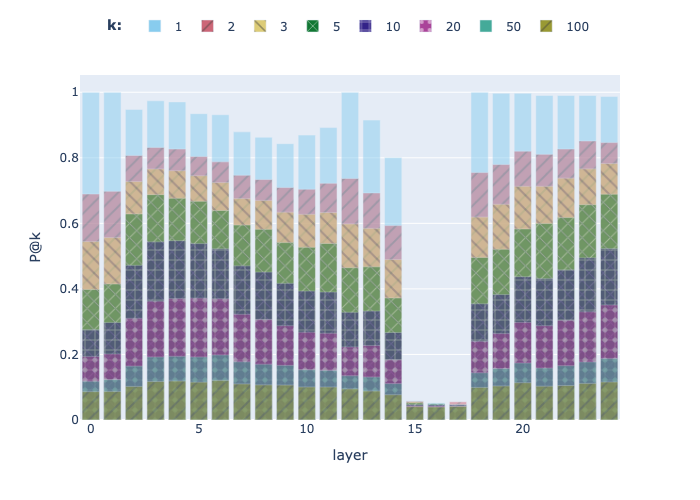}
    \caption{Word Representation}
    \end{subfigure}
    \hfill
    \begin{subfigure}{0.45\textwidth}
    \includegraphics[width=\linewidth]{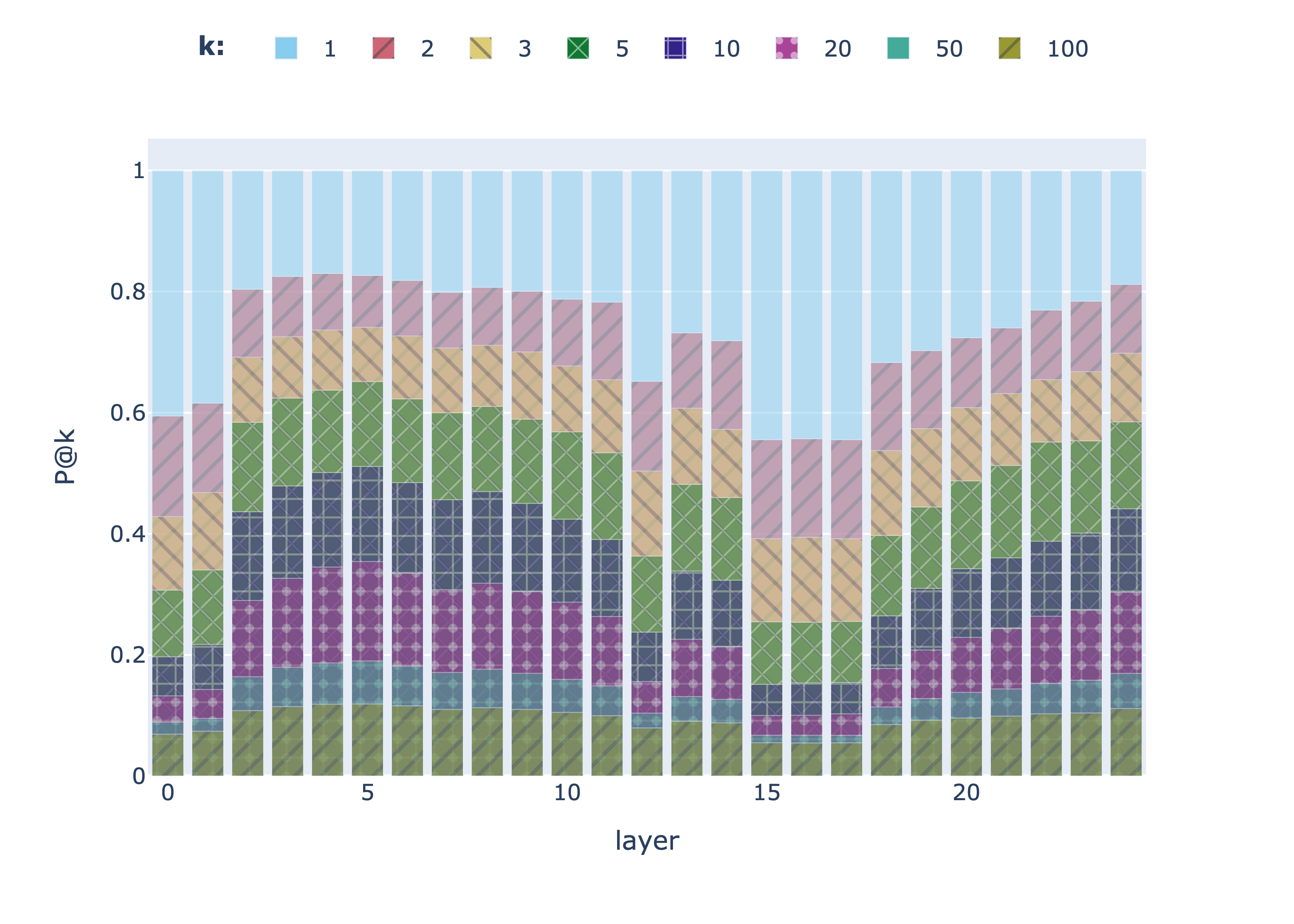}
    \caption{Contextual Representation}
    \end{subfigure}
    \caption{Precision@k as a function of the layer and neighborhood size $k$, calculated using the cosine distance. Each bar represents a specific layer, with stacked segments corresponding to different values of $k$.\label{fig:precision}}
\end{figure*}

\paragraph*{Results} The ability of \texttt{XLSR-53} contextual and word representations to retrieve words in corpora, measured by \texttt{Precision@k}, is illustrated in Figure~\ref{fig:precision}. Table~\ref{tab:full_res} then compares the precision and recall obtained with \texttt{MFCC} and \texttt{XLSR-53} representations for different values of $k$. In each case, we select the layer that achieves the best performance, assuming the optimal choice is known.\footnote{Our conclusions remain consistent when averaging scores across all layers, even if we find this operation more difficult to justify.}

\begin{table}
\centering
\begin{tabular}{lrrcrr}
\toprule
      & \multicolumn{2}{c}{Precision} && \multicolumn{2}{c}{Recall} \\
\cline{2-3} \cline{5-6}
$k$    &  \texttt{MFCC} & \texttt{XLSR-53} && \texttt{MFCC} & \texttt{XLSR-53}  \\
\midrule
 1   & 100.0\% & 100.0\% & & 5.7\% &  5.7\% \\
 2   & 57.2\%  &  85.1\% & & 6.4\% &  9.6\% \\
 3   & 40.4\%  &  78.3\% & & 6.8\% & 13.5\% \\
 5   & 27.5\%  &  68.9\% & & 7.5\% & 19.8\%\\
 10  & 16.8\%  &  54.7\% & & 8.8\% & 31.2\%\\
 20  & 11.5\%  &  37.2\% & & 11.6\% & 42.5\%\\
 50  & 8.1\%   &  19.9\% & & 19.3\% & 55.0\%\\
 100 & 6.4\%   &  12.1\% & & 29.8\% & 64.8\%\\
\bottomrule
\end{tabular}
\caption{\texttt{Precision@k} and \texttt{Recall@k} for the different representations we are considering. For each value of $k$ we selected the layer with the highest F\textsubscript{1} score.\label{tab:full_res}}
\end{table}

Several key observations can be made from these results. First, despite the anisotropy of the representations, we observe that the distance between words, as measured using DTW, has a meaningful interpretation. It is possible to retrieve words based on their raw representations by considering only the distances between embeddings, without additional training and/or fine-tuning. this result is also true for the last representations of the last layers, for which the anisotropy is strongest and the variance of cosine similarities almost zero. It also appears that the performance varies significantly across layers. The drop in performance is particularly marked on layers 15, 16 and 17, for which we have been unable to find any particular explanation. This observation emphasizes the importance of carefully selecting the appropriate layer for real-world applications, a conclusion that has already been drawn e.g., in \cite{bordes2023guillotine}. 

Figure~\ref{fig:precision} clearly highlights the impact of context in \texttt{XLSR-53} representations. When computing word representations (without incorporating contextual information), the precision at a neighborhood size of 1 is not always equal to 1. This contrasts with contextualized representations, where the nearest recording is consistently the one from which the query was extracted. This observation underscores the crucial role of context in \texttt{XLSR-53} representations. In our corpus, target sentence representations are always computed within context, whereas query word representations are not necessarily contextualized (as is the case in our word-level representations). This discrepancy introduces a mismatch, which explains why word representations perform worse than fully contextualized representations.


As expected, performance decreases as the neighborhood size increases, regardless of the layer considered. However, when focusing on immediate neighbors, we see that precision remains high: In a neighborhood of size 10, the precision is larger than 45\,\% for more than half of the layers. This suggests that, in the audio embedding space, representations of nearby words are often closely aligned. On the other hand, it is worth noting that even for larger neighborhoods ($k = 50$ or $k = 100$), recall remains low (recall that there are only 10 recordings of the same word) and does not exceed 65\,\%. This indicates that \texttt{XLSR-53} representations still exhibit significant variability, meaning that not all variants of the same word are closely clustered.

When comparing \texttt{XLSR-53} representations to \texttt{MFCC}, we observe that although precision decreases with $k$ for both representations, the decline is much slower for \texttt{XLSR-53} than for \texttt{MFCC}. To confirm this observation, we have included in the supplementary material the distance distributions between two utterances of the same word, using the three different representations we considered. By examining how these distributions evolve across layers, we observe that the distance between two occurrences of the same word decreases as the layers deepen, along with a corresponding reduction in variance. The comparison of \texttt{Precision@1} seems to show that this reduction captures mainly linguistic information.

All our observations suggest that \texttt{XLSR-53} representations create more structured spaces, where identical words are more closely aligned. This implies that \texttt{XLSR-53} representations are more abstract, as they build more robust representations that are less dependent on the audio signal content. Instead, they seem to capture more linguistic information, going beyond physical factors like the speaker’s voice, microphone, or the acoustic conditions of the recording environment.

\section{Conclusion}

Our study demonstrates that despite the strong anisotropy of \texttt{wav2vec2} representations, their similarity measures remain effective for keyword spotting without transcription. This suggests that these representations capture high-level phonetic structures that generalize well across speakers, making them robust for computational documentary linguistics. More broadly, our findings highlight the importance of pretraining in learning rich and invariant speech representations, which can benefit various speech processing applications.

\ifinterspeechfinal

\section{Acknowledgements}
This research was partially funded by the \textsc{DiagnoSTIC} project  supported by the \textit{Agence d'Innovation de Défense} (grant n\textsuperscript{o} 2022 65 007) and the \textsc{DeepTypo} project supported by the \textit{Agence Nationale de la Recherche} (ANR-23-CE38-0003-01).
\fi

\bibliographystyle{IEEEtran}
\bibliography{mybib}

\end{document}